\documentclass[10pt]{bmc_article}

\usepackage{cite} 
\usepackage{url}  
\usepackage{ifthen}  
\usepackage{multicol}   
\usepackage[utf8]{inputenc} 
\usepackage{color}

\usepackage{textcomp,graphicx,epsfig}

\newboolean{publ}

\newenvironment{bmcformat}{\baselineskip20pt\sloppy\setboolean{publ}{false}}{\baselineskip20pt\sloppy}

\begin{document}
\begin{bmcformat}


\title{Statistical data mining for symbol
       associations in genomic databases}
 

\author{Bernard Ycart\correspondingauthor$^{1,2,3}$%
         \email{Bernard Ycart\correspondingauthor - Bernard.Ycart@imag.fr}
       \and 
         Fr\'ed\'eric Pont$^{3,4,5,6}$%
         \email{Fr\'ed\'eric Pont - Frederic.Pont@inserm.fr}%
       \and 
         Jean-Jacques Fourni\'e$^{3,4,5,6}$%
         \email{Jean-Jacques Fourni\'e - Jean-Jacques.Fournie@inserm.fr}%
      }


\address{%
\iid(1) Universit\'e Grenoble Alpes, France\\
\iid(2) Laboratoire Jean Kuntzmann, CNRS UMR5224, Grenoble, France\\
\iid(3) Laboratoire d’Excellence 'TOUCAN', France\\
\iid(4) INSERM UMR1037-Cancer Research Center of Toulouse, Toulouse, France\\ 
\iid(5) Université Toulouse III Paul-Sabatier, Toulouse, France\\
\iid(6) ERL 5294 CNRS, Toulouse, France%
}%

\maketitle


\begin{abstract}
A methodology is proposed to automatically detect 
significant symbol
associations in genomic databases.
A new statistical test is proposed to assess the significance of a group of
symbols when found in several genesets of a given
database. Applied to symbol pairs, the thresholded
p-values of the test
define a graph structure on the set of symbols. The cliques of that
graph are significant symbol associations, linked to a
set of genesets where they can be found. The method can be applied to
any database, and is illustrated MSigDB C2 database. 
Many
of the symbol associations detected in C2 or in 
non-specific selections did correspond to already known interactions. 
On more specific selections of C2, many 
previously unkown symbol associations have been
detected. These associations unveal new candidates for
  gene or protein interactions, needing further investigation
  for biological evidence. 
\end{abstract}

\ifthenelse{\boolean{publ}}{\begin{multicols}{2}}{}


\section*{Background}
Large-scale genomic databases have been developed for over a decade 
as catalogs of genesets\cite{Schaefer04,Caryetal05}; a geneset is a
list of genes/proteins, the expression level of
which was found to be associated to some biological process, cellular
component, metabolic function, type of cancer, etc. Examples include
the KEGG database \cite{Kanehishaetal04}, MSigDB C2 to C7 
\cite{Subramanianetal05}, and those from the Gene Ontology Project
\cite{GO00}. The contents of genesets will be viewed here as
  \emph{symbols} i.e. character strings without a priori biological
  meaning, and will be distinguished from the gene or proteins they represent.

Parallel to the creation of databases, the question
of building a full network of Protein-Protein Interactions (PPIs), or
interactome, has
also received a lot of attention
\cite{JonesThornton96,RivasFontanillo10,Vidaletal11,Kirouacetal12}. 
De Las Rivas and Fontanillo
\cite{RivasFontanillo10} 
distinguish binary methods that look for pairwise associations, from
co-complex methods that detect groups of more than two proteins. One seemingly
simple co-complex method consists in listing geneset intersections
in databases. Indeed, finding a given
group of symbols in several genesets provides a reasonable heuristic for a
possible PPI, to be validated by a subsequent biological study.

Our goal here is not to discuss the biological relevance of co-complex
PPIs, but instead to propose a new methodology to 
automatically detect in genomic databases the presence of such
associations, without any prior biological knowledge.
Our question is: how can symbol groups being present in a significant
number of different genesets be \emph{systematically}
detected in a given database?
The answer seems straightforward: in theory,
it should suffice to list all possible geneset intersections 2 by 2, 
then 3 by 3, etc. The difficulty
comes from combinatorial explosion: if there are $p$
genesets, the number of geneset intersections is $2^p-p-1$, i.e. 
$2.9\times 10^{1421}$ for the $4722$ genesets of MSigDB C2. 
The list of \emph{all} geneset intersections will
remain forever out of reach. Systematically finding
sizeable intersections in a given collection of sets has long been one
of the main problems of datamining, since the introduction of the
first frequent itemset algorithms by Agrawal et al \cite{Agrawaletal93}: see
\cite{Goethals05,Hanetal07,Borgelt12} for general reviews,
\cite{deGraafetal05} for an application in the context of genomic
profiling. However, we argue that algorithms that 
systematically detect the most frequent sets are not adapted 
to the present context. Indeed, the
most frequent associations involve symbols present in many different 
genesets: their associations are the most conspicuous
and well documented. Associations of relatively
unfrequent symbols are potentially more interesting. This poses the
problem of assessing the
significance of a given intersection.

Contrarily to existing algorithms such as that of Kirouac et
al. \cite{Kirouacetal12}, the method proposed here relies on a
purely statistical approach.
A new test has been defined: it computes a p-value for each
symbol group common to a given collection of genesets. 
The test takes
into account the frequencies of the different symbols in the
database: an association
of frequent symbols  is less significant than an association of
rare symbols in the same number of genesets. Considering a given
group of symbols, the test 
statistic is the number of genesets it appears in. Under the null
hypothesis of random occurrences, its distribution can be
approximated by a Poisson distribution, using classical results on the
so called ``law of small numbers'' \cite{Janson94}. The p-value
is computed as a tail probability of the Poisson approximation.

The test is used to define an undirected weighted graph
structure for which vertices are 
symbols: each pair of symbols (edge) is
weighted by its p-value in the 
association test. Thus p-values are viewed as distances
between symbols: the more frequent the occurrence of the
pair, the lower 
the p-value, and the
closer the two symbols. Using this type of mathematical structure in
genomics data 
mining is not new: see the review by Lee et
al. \cite{Leeetal08}. Once the weights have been calculated, two
procedures can be applied. One of them  uses pairwise p-values as
a dissimilarity to perform a hierarchical clustering of the set of
symbols under consideration
\cite{Eisenetal98,HussainHazarika11}. The other 
consists in thresholding the weights to deduce a
continuum of unweighted graphs on the set of symbols: 
a threshold $h$ being chosen, an edge
exists between two symbols if the p-value of the pair is smaller than
$h$. It is then natural to consider as significant 
the \emph{maximal cliques} of the
thresholded graph \cite{ButenkoWilhelm05}. Indeed, 
cliques are groups of symbols
such that any two of them
are connected in the graph, { which is equivalent to
  saying that any two of them are
associated in a significant number of genesets}. 
Algorithmic complexity is a major
difficulty here. Clique finding is a NP-hard problem, and listing
all cliques of a reasonably dense graph is not feasible in practice
beyond a few hundred vertices. In our case, the possibility to adjust the
threshold is a crucial feature. The lower the threshold $h$, the
sparser the graph. For a given database, $h$ can be chosen
such that the number of neighbors of each symbol (its degree in the
graph) is smaller than $100$, say. For such a sparse graph, 
the classical Bron-Kerbosch algorithm can be
applied to the neighborood of each symbol \cite{Harleyetal01}. This
yields a list of all cliques of the graph. Each clique (significant
group of symbols) is then examined to see if it appears
in two or more 
genesets, then possibly completed by other symbols
appearing in the same 
genesets; lastly, the association p-value of the whole group is computed.

The procedure has been implemented in a R script \cite{R} available
online together with files of results.
Examples of executions on MSigDB C2 
  \cite{Subramanianetal05}
(referred to as \texttt{C2} thereafter)
and five  selections from the same  are given: 
see next section and additional files. When genesets cover many
diseases or functions, such as those of \texttt{C2}, or even
non-specific selections like
all cancer-related genesets of \texttt{C2},
the majority of detected associations are compatible
with known PPIs. In our view, this supports the intuition 
that significantly large geneset intersections do contain PPI information. On 
specific selections of \texttt{C2}, such as for instance genesets
related to breast cancer, a majority of detected associations did not
correspond to known PPIs. Further investigation should be made to
assess their biochemical signification.
We are aware that an algorithmic listing of
significant associations does
not necessarily imply that all listed groups correspond to
meaningful PPIs. Such a listing must necessarily be expert-curated
for biochemical validation. We are also aware that our specific
selections of genesets are too restrictive to be completely meaningfull.
They are presented here only as an illustration of potential uses, 
hoping that the method could prove useful to 
disease-oriented interactomics \cite{Vidaletal11}.
\section*{Results}
Results of the in silico experiments that were conducted to assess the
interest of the method are reported here. Two points had to be
proved. The first point was that the information contained in large geneset
intersections was compatible with established interactome
knowledge. The second point was that significant geneset intersections
such as detected by our method
could contain previously unkown PPI information. In order to establish
those two points, we have applied the method to
  \texttt{C2}, and five other 
  databases, obtained by selecting in \texttt{C2} those genesets with
  names matching one ore more character strings. They will be referred
  to as \texttt{C2\_Blast} (character string ``blast''),
\texttt{C2\_Breast} (character string ``breast''), 
\texttt{C2\_Cancer} (any character string related to cancer, such as
``tumo'', ``carci'', etc.), 
\texttt{C2\_K} (character string ``KEGG''),
\texttt{C2\_Lymph} (character strings ``lymph'' or ``leuk'').

To begin with, the application of the method to the
  different databases will be described.
As a first step, the association graph must be explored by
  examining the degrees of symbols at different
  thresholds. A plot of degrees
  against frequencies at a given threshold illustrates the sparsity of
the graph. Such a plot appears on Figure 1: all
symbols in \texttt{C2\_K} have been plotted by their
frequency on the x-axis, and their degree on the y-axis 
at threshold $h=0.001$. Figure 2 shows the same plot on the full
database \texttt{C2} at a much lower threshold $h=10^{-15}$. Very
frequent symbols do not necessarily have the
highest degrees, since p-values for associations involving frequent
symbols tend to be larger (see the ``Association test''
  section).

Table 1 shows, 
for a selection of twelve symbols, the
frequencies, and the degrees at thresholds
$10^{-2},\ldots,10^{-10}$ in the full database \texttt{C2}. 
Observe that COMP, which is present in 35 genesets of \texttt{C2}, has many
more neighbors at threshold $10^{-2}$ than MAPK1 which is more
frequent. On the contrary, at threshold $10^{-6}$, MAPK1
still has 21 neighbors, whereas COMP has only 8. The way the
association test has been designed decreases the number of neighbors
of very frequent symbols for low values of $h$. However high numbers
of co-occurrences (as in the case of MAPK1 with some of its neighbors)
are still translated into 
very low p-values. It is often helpful to vizualize local parts of the
graph. For the same symbols as in Table 1, Figure
3 shows  
two neighborhoods in the association graph, at the same threshold $10^{-6}$.

The choice of a threshold $h$ is left to he user. For statistical
reasons, a significance threshold larger
than $5\%$ is not appropriate. For algorithmic reasons,
a threshold such that the highest degree in
the association graph is about $100$ should be selected if possible.
For each of the five databases, a threshold was chosen and the
detection algorithm 
was run. The thresholds that have been applied to the six examples 
are shown on Table 2. The six lists of symbol associations are given
as additional text files. The larger the number of
genesets, the denser the association graph at a given threshold, thus
the lower the threshold should be. 
The number of symbols and the number of genesets of each association
varies. Figure 4 shows a scatterplot of both
quantities for the 828 associations detected at threshold $10^{-5}$ in
\texttt{C2\_Cancer}. Similar plots were obtained on all
databases. Associations represented on the bottom right corner correspond
to large numbers of symbols common to few genesets. 
As an example coming from \texttt{C2\_Cancer}, 
the two genesets ``Acevedo liver cancer
up'' (973
symbols) and ``Acevedo liver tumor vs normal adjacent
tissue up'' (863 symbols) 
have 494 symbols in common, a high rate of overlap
indicating \emph{informational redundancy} (the two geneset definitions are
almost synonymous).       
This phenomenon is common to all
databases: some of the associations detected by our method 
are very large groups of
symbols, common to a small number genesets.  We believe
that such large intersections should be interpreted with caution, as
the largest overlaps are most likely to result from direct
informational redundancy between genesets rather than actual
biological associations.

On the contrary, associations in the top left corner
  of Figure 4 involve fewer symbols common to many genesets. These symbols
  usually correspond to very common `jack-knife' proteins 
(AKTs, COLs, ERKs, MAPKs, \ldots) 
involved in many different cell functions and biological pathways. 
As an example, the highest two points on Figure 4 correspond to pairs of
collagens:
COL1A1, COL1A2 found together in 25 genesets, 
COL1A2, COL3A1 found together in 23 genesets. Actually, the three
collagens COL1A1,
COL1A2, COL3A1 are found together in 43 genesets of \texttt{C2}. 
This can hardly be
considered as new biological information, but rather as \emph{biological
redundancy}. We shall argue in the
discussion section that finding together symbols of the same family
in some pathways may be biologically interesting, though not surprising.

Here is an example of both informational and biological
redundancy. LOC652826 is present in 47 genesets of \texttt{C2}
(1151 symbols out of the 21047 of
\texttt{C2} are locations). Out of those
47 genesets, 46 have a name beginning with ``reactome''. The intersection
of those 47 genesets is made of LOC652826, PSMC6 and 36 other PSMs
(proteasomes). Actually, LOC652826 is a synonym of
PSMC6. Nevertheless, the pair LOC652826, PSMC6 is detected as significantly
associated by our test (P$=5.8\times10^{-38}$) and the intersection of
the 47 genesets is identified as a significant association by
the algorithm.

Both types of redundancy will be further illustrated by the following
associations detected in \texttt{C2\_K}:
AKT1, AKT2, and AKT3 found together in 30 genesets;
MAPK1 and MAPK3 found together in 46 genesets;
PIK3CA, PIK3CB, PIK3CD, PIK3CG found together in 34 genesets;
AKT1,   AKT2,   AKT3,   GRB2,   HRAS,   MAP2K1, MAPK1,  MAPK3, 
PIK3CA, PIK3CB, PIK3CD, PIK3CG, PIK3R1, PIK3R2, PIK3R3, PIK3R5,
RAF1,   SOS1,   SOS2 found together in 16 genesets.
It is not
suprising to see homologs 1,2,3
of v-akt murine thymoma viral oncogenes (AKTs) jointly appear in
\texttt{C2\_K} genesets. The same can be said of mitogen-activated
protein kinases (MAPKs) and phosphoinositide-3-kinases (PI3Ks). 
The largest association found in \texttt{C2\_K} involved 95
symbols, including 
22 IFNs (interferons) and 47 ILs (interleukins). They were common to
two genesets,
``Cytokine cytokine receptorinteraction''
``Jak stat signaling geneset''. 
{This corresponds to a typical case of
informational redundancy as most cytokine/cytokine receptor
interactions trigger intracellular signals which 
are transduced through Jak/Stat cascades.} 
The second largest association was that of 90 symbols, including
14 ATPs (ATP synthases, H+ transporting), 20 COXs (cytochrome c
oxidases),
35 NDUFs (NADH dehydrogenase (ubiquinone)), 4 SDHs (succinate
dehydrogenase complex), and 8 UQCRs (ubiquinol-cytochrome c reductase).
Those 90 symbols were found in 4 genesets, named 
``Alzheimers  disease'', ``Huntingtons disease'',
``Parkinsons disease'', and ``Oxidative phosphorylation''.
{By contrast, this
association has a higher informative biological significance. 
The three distinct neurodegenerative diseases do involve neuronal
apoptosis, wherein a key step is defective mitochondial respiration,
also known as oxidative phosphorylation.

Here is a much less impressive looking association, still detected in
\texttt{C2\_K}:} 
COMP, THBS1, THBS2, THBS3, THBS4 found together in 3 genesets.
It concerns
relatively unfrequent proteins: THBS1 appears in 5 genesets, whereas
COMP, THBS2, THBS3, and THBS4 do not appear in any other than the 3
genesets they all have in
common. 
Among other interactome databases, we have chosen 
STRING 9.0 \cite{Szklarczyketal09} as a reference, and systematically
compared {symbol associations} 
detected by our method to STRING evidence views. 
The cartilage oligomeric matrix protein (COMP)
is not signaled in STRING as 
biochemically linked with thrombospondins (THBSs). 
Yet, finding them together does have a biological interest.
Indeed,
COMP is not a thrombospondin, yet a close examination of its structure
and functions evidences a link not detected by current algorithms or search
robots: the COMP includes a thrombospondin-like
domain.

It can be considered that the potentially novel associations
are likely to be found among those with a small enough number of symbols, and a
large enough number of genesets. Thus the lists can be screened over
numerical criteria. An example of screening (number of symbols
smaller than 10, number of genesets larger than 2) 
appears on columns 4 to 6 of Table 2. After numerical screening, the
remaining associations were tested in STRING 9.0
\cite{Szklarczyketal09}. STRING
distinguishes evidence of association according to
neighborhood, gene fusion, cooccurrence, coexpression, experiments,
databases, textmining, homology. We considered that two symbols were 
connected in STRING if at least 
one of the 8 links exists,
i.e. if there exists at least one edge in the evidence view. 
The results only reflect the
status at the date when comparisons were made. STRING is in constant
evolution, and includes new interactions almost daily. Several of the groups found
disconnected when the comparison was made, may have been connected
since.

Among the associations detected in the full database 
\texttt{C2} at threshold $h=10^{-15}$, nearly all fell
  under informational and/or biological redundancy. Very few
  disconnected STRING graphs were detected in that experiment; 
examples include:
PRLHR, DRD5 found together in 12 genesets;
UQCRC1, SDHA found together in 23 genesets;
ZNF367, UHRF1 found in 24 genesets. In 
  \texttt{C2\_K} and \texttt{C2\_Cancer}, a majority of detected
  associations also corresponded to
  STRING-connected graphs. In the other three (more specific)
  selections, a majority corresponded to disconnected, or even empty
  graphs. Here are two examples of STRING-disconnected associations from
  \texttt{C2\_Breast} (many more can be found in the corresponding
  additional file):
ERBB3, MYB found together in 7 genesets;
DSC3, KRT14, PDZK1IP1 found together in 6 genesets. 
Once again, algorithmic
detection cannot be considered a proof that ERBB3  (v-erb-2 erythroblastic
leukemia viral oncogene homololog 3) and
MYB (v-myb myeloblastosis viral oncogene homolog) are functionally
related, even though it has been shown that
both genes are deregulated by mutations of the
transcription factor TWIST in human gastric cancer
\cite{Fengetal09}.

To conclude this section, we mention another possible use of
our statistical test.
Once the p-values of joint appearances have been calculated for all
pairs of symbols in a database, the matrix of p-values
so obtained can 
be used as a
matrix of dissimilarities to perform a hierarchical
clustering. Several clustering methods have been discussed at length in the
literature \cite{Eisenetal98,HussainHazarika11}. It can be checked that
clustering from the
p-value matrix usually yields clusters wich are coherent with already known
biological information, when available. As an example, consider the association 
AKT1,   AKT2,   AKT3,   GRB2,   HRAS,   MAP2K1, MAPK1,  MAPK3, 
PIK3CA, PIK3CB, PIK3CD, PIK3CG, PIK3R1, PIK3R2, PIK3R3, PIK3R5,
RAF1,   SOS1,   SOS2 found in 16 genesets of \texttt{C2\_K}. Figure 5 
shows a hierarchical
clustering obtained through the single link algorithm: the clusters
match known PPIs, and groups of proteins with the strongest
biochemical relation are correctly identified as homogeneous
clusters. 

%
\section*{Discussion}
As for many other datamining tools, the objective of our method is to
algorithmically reduce combinatorial explosion in searching for
sizeable intersections from collections of genesets. Unlike existing
frequent itemset searching algorithms
\cite{Goethals05,Hanetal07,Borgelt12,deGraafetal05}, the goal here is not
to list all groups of at least so many sets intersecting
in so many items. This would output only very frequent
symbols, excluding all others. Relatively small
intersections of rather unfrequent symbols may be much more
significant, and therefore should be enhanced. This is done by
selecting intersections on their p-value from a statistical test, rather
than on their sizes. Unlike in \cite{Kirouacetal12}, the association
graph that is deduced from testing pairs of symbols 
is not meant as the representation of an interaction
network, but as the basis of a co-complex PPI searching method: 
the cliques of the graph should be tested as possible interaction
candidates. The method can be applied to any database, and we believe
that reducing generic databases to more specific genesets can lead to
interesting lists of associations, small enough to be expert
curated. Indeed, an algorithmically detected association cannot be
accepted as biological evidence, but only as a possible candidate,
selected on statistical evidence.

Examining the examples of outputs given as additional files, one cannot fail
to notice very large intersections of several genesets under similar
names. 
 For different reasons, linked to the way they were compiled,
 largely overlapping genesets 
have been included
in most databases: this can be called `informational
redundancy'. It can be controlled by algorithmically screening outputs
on numerical criteria (e.g. eliminating too large
intersections). Also, among detected intersections, many include genes
belonging to the same family. The definition of our statistical test
enhances significant
associations of relatively unfrequent symbols, but does not remove
conspicuous associations of very frequent symbols.
Previously given examples include AKTs,
COLs, MAPKs, PIKs, PSMs, etc. This is part of what could be called `biological
redundancy'. Indeed it is not suprising to observe two members of the
same family
jointly appear in many different genesets. But is it completely
uninteresting? We do not believe so and detail several arguments
below.

Associations involving redundant genes might define a
functional group which is highly informative. As an example consider
the `redundant' association 
CD1A, CD1B, CD1C, CD1D, CD1E. These five structurally
related glycoproteins have almost similar
functions. But for biologists, `almost similar' very
often means `actually distinct'. Indeed,
these five CD1s do mediate seemingly similar but very
distinct immunological functions, related altogether to cell surface
presentation of non-peptide antigens to T lymphocytes. In short, CD1A
presents a group of mycobaterial glycans, CD1B presents lipids, CD1C
presents glycolipids and sulfatides, CD1D presents a ceramide and CD1E
does not presents antigens but processes cytoplasmic
phospholipids. Hence an association comprising several of the redundant CD1s
is informative of a more global process involving immunity to
non protein antigens (the so-called innate immunity). In addition, 
finding other non-CD1 genes in the same association is quite
interesting for a biologist. It turns out that in \texttt{C2}, two
genesets contain the five CD1s above, and also MME (membrane
metallo-endopeptidase) and DNTT (deoxynucleotidyltransferase,
terminal).

We do not believe that associations involving functionally redundant
genes necessary lack interest. The presence of
`redundant' genes in some detected association does bring an information 
to the reader, as globally `redundant genes' are never exactly
redundant, due to evolutive speciation. In any metazoan genome,
duplicated genes with the 
        same initial function always evolve separately, and
        progressively acquire more specific functions along evolution,
        time, and selective pressure at the genic level. Hence
        `functional redundancy' does not mean `non-novel'. Four
        examples will be given in different fields of biology.

\emph{Example 1 (evolutionary developmental biology):} 
One could quote the very first Toll genes which
still control development in the fly, but were ancestrally duplicated
several times leading to ten Toll-Like Receptor genes in the human
genome (eleven in the mouse). None of these TLRs does exactly the same
as in the original fly: roughly summarizing, they all
control inflammation in mammals, although actually none of them does
the same thing in this function (see Beutler and Hoffmann's work on
the activation of innate immunity that won them the 2001 Nobel prize
in medicine). The 
same would apply to the conserved and functionally redundant NLR
genes. Hence TLRs-comprising associations can be seen as biologically
redundant but nevertheless have a different significance according to
which TLR members are included.  

\emph{Example 2 (immunology):} Consider two genes from the large subgroup of
KLRs: KLRC1 and KLRK1. These two 
genes mediate the same ligand recognition in the
same immune process (regulation of NK cell-mediated lysis of target
cells), but they actually transduce opposed signals:
KLRK1 activates cell lysis while KLRC1 inhibits it. Their dual
presence in a clique would not mean the same thing (regulation of NK
activity) than the presence of several of the redundant KLRK-like
genes only (activation or inhibition). Actually, KLRC1 and KLRK1
are both present in seven different genesets of \texttt{C2}, together with
KLRD1.

\emph{Example 3 (neurobiology):} Adenosine (dopamine, 
or other neuromediator)
receptors expressed in animal brains, are functionally
redundant: they encode for a vital receptor that receives and
transmits a signal which contributes to the local tissue homeostasis
and function. At the whole body scale, vital functions are fulfilled
by a normally functional brain. Consider the four redundant
adenosine receptor genes (ADORA1, ADORA2A, ADORA2B, and
ADORA3). Their common presence in six different genesets of \texttt{C2}
is somehow puzzling, as these four
genes are expressed in very different tissues (distinct promoter
sequences), and transduce differently. ADORA2A and ADORA2B are
coupled to Gs transducing genes, whereas the two others are
coupled, on the converse, to Gi transducing
genes. Moreover two ADORAs open ion channels, the two others raise
intracellular cAMP. Altogether, these four seemingly functionally
redundant adenosine receptor GPCR are actually not so redundant.

\emph{Example 4 (pharmacology):} Similar points could be made using 
many other examples. Most of the Ig's superfamily, of the
very large TNFR superfamily, of the GPCR superfamily he olfactory
receptor (OR) superfamily did evolve by serial
duplication/mutation/neofunctionalization and functional speciation
(among other processes) which led to the presently puzzling functional
redundancy in most of the eukaryote genomes. For example genes
encoding for the various isoforms of PI3K are grossly depicted as
functionally similar isoforms, yet each of these is involved in very
different signalling according to the tissues and cell types in which
they are expressed, (leading to their very selective targeting for the
therapy of specific cancers).

Frequent genes in databases correspond to  `jack-knife genes'
(MYCs, ERKs, MAPKs, PI3Ks...) which are involved in many different cell
functions and biological pathways. Indeed duplication of such genes
along evolution has led to gene families with all possible strengths
of penetrating phenotypes, from hypomorphics to functionally distinct
mutants. Undoubtedly on the long term, gene duplication also drove to
speciation of functions and functional divergence, as proposed long 
ago by S. Ohno \cite{Ohno70}. 
Multifunctional genes evolving from an ancient
unique function to multiple neofunctionalization, by various possible
evolutive processes (see \cite{ConantWolfe08,DittmarLiberles10}) are numerous in
our genome and represent a major source of `functionally redundant'
associations. By contrast, the currently monofunctional genes present
duplications and appear in `functionally redundant' associations far less
frequently. This observation
is not trivial, and links blatantly the significance of biologically
redundant associations to molecular evolution (see
e.g. \cite{HittingerCarroll07}).  

\section*{Conclusions}
We have defined an algorithm that automatically outputs
symbol 
associations by searching for significant geneset intersections
in a database. The method is based on a statistical test, used
to define a graph structure among symbols. It has been
applied to MsigDB C2 \cite{Subramanianetal05}
and five databases selected from the same; two of
them had little specificity (\texttt{C2\_K},
\texttt{C2\_Cancer}), the three others were more specific 
(\texttt{C2\_Apop}, \texttt{C2\_Blast},
\texttt{C2\_Breast}, \texttt{C2\_Lymph}). On each database, a list of
symbol associations,
small enough to be expert-curated, was obtained.  The detected associations 
were compared to STRING evidence views. Out of the associations coming from 
non-specific databases, a
majority had connected graphs in STRING evidence views; this validates
the intuitive idea that significant geneset intersections
correspond to biologically relevant interactome information. Among
specific databases, many detected
associations had disconnected STRING graphs; this may be
an indication that new interactome information can be
extracted. Therefore, we believe that the proposed 
method can be added to the data mining tools for searching protein-protein
interactions.
\section*{Methods}
\subsection*{Association test}
The association test will be defined in this section.
A basic assumption is that there are no duplicates in genesets: each
symbol appears at most once in a given
geneset. The
distinction between symbol and protein or gene is crucial:
it is sometimes the case that two symbols actually
correspond to the same gene, even though they are treated here as
different (an example has been given in
the Results section). 

Consider a database made of $p$ genesets of different sizes:
the $i$-th geneset contains $l_i$ symbols, assumed to be all
distinct. Denote by $n_j$ the frequency of symbol number $j$, i.e.
the total number of genesets it appears in.
The
assumption of no duplicates implies that the sum of symbol 
frequencies is equal to that of geneset sizes: $\sum_i
l_i=\sum_j n_j$. Let us denote by
$N$ that sum: $N$ is the total number of symbol occurrences in the
database. The null hypothesis
of our test (lack of information) is that the genesets have been
constituted by independently including the different
symbols. 
Under that null
hypothesis, the probability that symbol number $j$ appears
in geneset number $i$ can be estimated by $(n_j\times
l_i)/N$. Consider a set of $k$ different
symbols, labelled $j_1,\ldots,j_k$. If the
appearances are assumed to be independent, the
probability that the $k$ symbols are found together in
geneset $i$ must be the 
product:
$$
p_i= \frac{n_{j_1}\times l_i}{N}\cdots \frac{n_{j_k}\times l_i}{N}
=
\left(\frac{l_i}{N}\right)^k \left(n_{j_1}\cdots n_{j_k}\right)\;.
$$
The total number of genesets where the group can be found is the sum
over all genesets, of independent Bernoulli random variables with
parameter $p_i$: the $i$-th random variable is $1$ if the group is
present in the $i$-th geneset (which occurs with probability
$p_i$), $0$ else. For $k$ large enough and even for large genesets, 
the probabilities
$p_i$ are small. By the law of small numbers \cite{Janson94}, the
distribution of the sum of a large number of Bernoulli random variables with
small parameters can be approximated by a Poisson distribution, the
parameter of which is the sum of all probabilities. The sum of all
$p_i$'s can be
interpreted here as the expected number of genesets the group should be
found into, if symbol appearances were independent. 
Let us denote it by $\lambda$.
\begin{equation}
\label{lambda}
\lambda = \sum_i p_i = \left(n_{j_1}\cdots n_{j_k}\right)
\sum_i \left(\frac{l_i}{N}\right)^k \;.
\end{equation}
Assume now that the group of symbols $j_1,\ldots,j_k$ has been found
in $x$ different genesets. The p-value associated to this observation 
is the right tail probability at $x$ (probability to be larger or
equal to $x$) for the Poisson distribution with parameter $\lambda$. 
Observe that $\lambda$ is proportional to each
  $n_{j_h}$; so when the frequency of symbols increases, 
  $\lambda$ increases too, and so does
the tail probability for a given number of occurrences. This is why
associations of very frequent
  symbols are considered less significant by the test.

In our method, p-values must be calculated for each pair of
symbols in the database, which may seem prohibitive as
far as computing 
time is concerned. Let us say that the p-values of all pairs
containing symbol 
number $j$ must be calculated: here is how the algorithm has been implemented.
Firstly, the geneset sizes $l_i$, the symbol frequencies $n_j$, the
total number of occurrences $N$, and the sum of squares 
$\sum \left(\frac{l_i}{N}\right)^2$ are precalculated with negligible
cost, using the function \texttt{table} of R. 
Then the database is reduced to those genesets containing
symbol number $j$, making it much smaller. 
The reduced database is then analyzed: the
symbols it contains are those which can be found together with
symbol number $j$. The number of times they occur in the reduced
database is the number of joint occurrences of the corresponding
pair in the full database. The function
\texttt{table} outputs a
table of joint frequencies of pairs, labelled with those symbols
paired with symbol number $j$. Using the precalculations, a
table for the corresponding Poisson parameters $\lambda$ is made
and 
Poisson tail probabilities is
calculated at low computing cost.
This has been implemented in the function  \texttt{neighbor.symbols} 
from the \texttt{spa.r} script available online. That function is
repeatedly applied to all symbols in the function
\texttt{database.graph}. Even for the largest databases available to us,
its total execution time is of the order of the hour on a PC. The results
can be repeatedly used for different graph structures, as will
be explained in the next section. They can be automatically saved
as a R data file and recalled for future use.
\subsection*{Association graph}
\label{graph}
Once all pairwise p-values have been calculated, they are viewed as
a weighted graph structure, symbols being taken as vertices of the
graph. Observe that if two symbols cannot be found together in any
geneset, the corresponding pairwise p-value is $1$. Two symbols
with a small pairwise p-value can be seen as neighbors: 
the smaller the
p-value, the closer the neighbors.

Let $h$ be chosen,
positive and smaller than $0.05$. Those pairwise p-values smaller than
$h$ define an undirected graph, called the association graph at
threshold $h$. In that graph, two symbols are joined by an edge 
if the number of genesets were the pair can be found is
significantly high at threshold $h$. The number of neighbors of a
symbol, i.e. its degree in the graph, decreases as the threshold
decreases.

For a small threshold $h$, it is natural to consider the
cliques of the association graph at threshold $h$, i.e. 
groups of pairwise connected vertices
\cite{ButenkoWilhelm05,Harleyetal01}. Any clique containing a given
symbol, is necessarily included in the set of neighbors of that
symbol. If the number of neighbors is relatively small (smaller than
100, say), then all maximal cliques in the set of neighbors can be
listed by the Bron-Kerbosch algorithm in reasonable computer time
\cite{Harleyetal01}. The \texttt{maximal.cliques} function 
of the R package \texttt{igraph} by Csardi and Nepusz \cite{Csardi06} was
used.

Maximal cliques can
then be tested to check whether they appear in a significant number of
genesets as a whole.  It is
natural to complete all detected maximal cliques
by those symbols appearing in the same genesets. Once all maximal
cliques have been detected and completed, duplicates are eliminated,
the association test is applied to each completed clique, and the results are
returned as a list (cf. additional files).
\section*{Competing interests}
The authors declare no competing interest.

\bigskip

\section*{Author's contributions}
BY conceived the statistical test, implemented the method and drafted
the manuscript. FP designed the parsing routines to extract genesets
as text files, helped to refine the R code, and to draft the
manuscript. JJF curated the outputs from STRING and helped to
draft the manuscript. All authors read
and approved the final manuscript.

\section*{Acknowledgements}
  \ifthenelse{\boolean{publ}}{\small}{}
  All authors acknowledge financial support from 
Laboratoire d'Excellence TOUCAN (Toulouse
  Cancer). They are indebted to Christophe Cazaux for the initial
  idea, and to Nadia Brauner for helpful discussions and references.
They are grateful to the anonymous reviewers for important
  suggestions. 
 

\newpage
{\ifthenelse{\boolean{publ}}{\footnotesize}{\small}
 \bibliographystyle{bmc_article}  
\bibliography{/home/ycart/recherche/Fournie/JJ.bib} } 


\ifthenelse{\boolean{publ}}{\end{multicols}}{}



\section*{Figures}
%
\subsection*{Figure 1 - Frequency and degree of symbols in 
\texttt{C2\_K}}
All symbols in \texttt{C2\_K} have been plotted by their
frequency on the x-axis, and their degree at threshold
$h=10^{-3}$. {For each frequency, the symbol with highest degree
appears in red.}
Very
frequent symbols, such as AKTs or MAPKs do not necessarily have the
highest degrees, since p-values for associations involving frequent
symbols tend to be larger.

%
\begin{figure}[!ht]
\centerline{
\includegraphics[width=15cm]{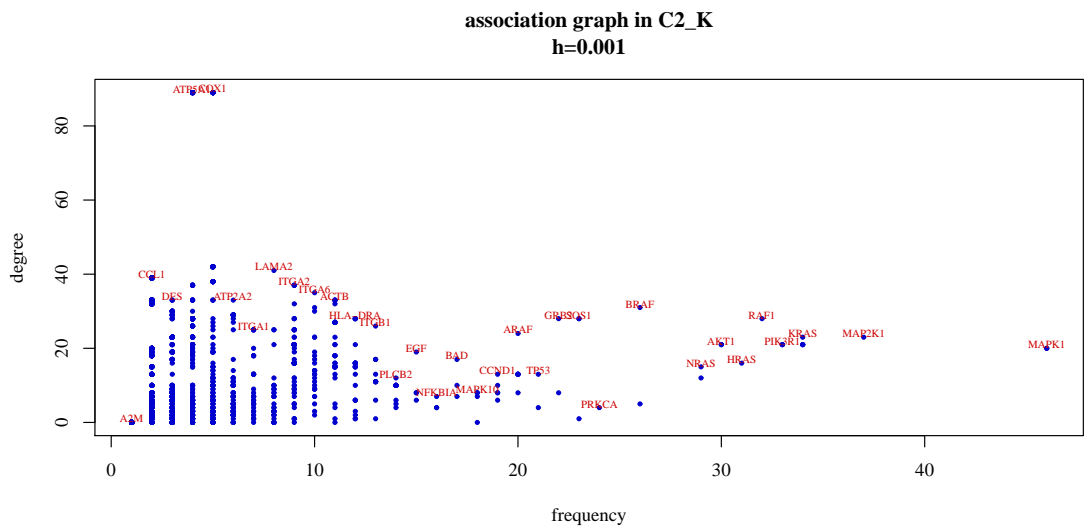}
} 
\caption{Frequencies and degrees of symbols in the \texttt{C2\_K}
  database. Each symbol is plotted by the number of genesets 
containing it (x-axis), the number of other symbols for which the
  pair p-value is smaller than 0.001 (y-axis).}
\label{fig:FDK}
\end{figure}
%

\subsection*{Figure 2 - Frequency and degree of symbols in 
\texttt{C2}}
All symbols in \texttt{C2} have been plotted by their
frequency on the x-axis, and their degree at threshold
$h=10^{-15}$. {For each frequency, the symbol with highest degree
appears in red.} 
%
\begin{figure}[!ht]
\centerline{
\includegraphics[width=15cm]{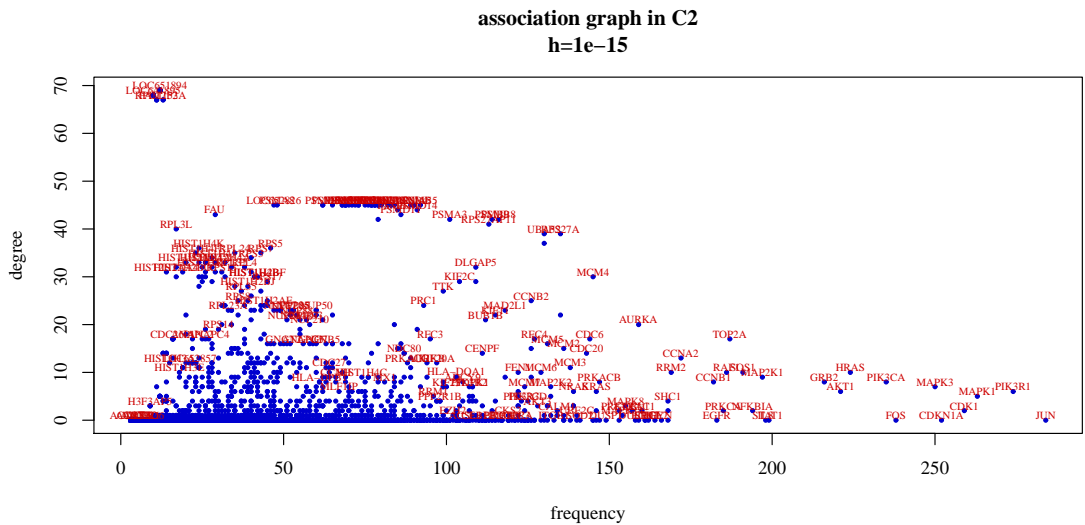}
} 
\caption{Frequencies and degrees of symbols in \texttt{C2}
  Each symbol is plotted by the number of genesets 
containing it (x-axis), the number of other symbols for which the
  pair p-value is smaller than $10^{-15}$ (y-axis).}
\label{fig:FDC2}
\end{figure}
%

%

\subsection*{Figure 3 - Association neighborhoods in 
\texttt{C2}}
Association graphs at threshold $10^{-6}$ in \texttt{C2} 
for symbols AKT1, GRB2, HRAS, MAPK1, PIK3CA, RAF1, SOS1 (left panel)
and  COMP, THBS1--4 (right panel) and their neighbors.
Consider for instance cartilage oligomeric
matrix protein (COMP). In STRING 9.0, it is mentionned as related to
collagens (COLs), SOX9, SLC26A2, HAPLN1, SDC1, ECM1. In \texttt{C2\_K}, 
among neighbors of COMP with p-value smaller than $0.01$,
THBs (thrombospondins) are the closest neighbors, collagens (COLs)
come next, together with CHAD, IBSP, RELN, TNs (tenascins), etc. 
Then come laminins (LAMs) and several others: the closest neighbors in the
full database may be different from those in a selection.
%
\begin{figure}[!ht]
\centerline{
\begin{tabular}{cc}
\includegraphics[width=7cm]{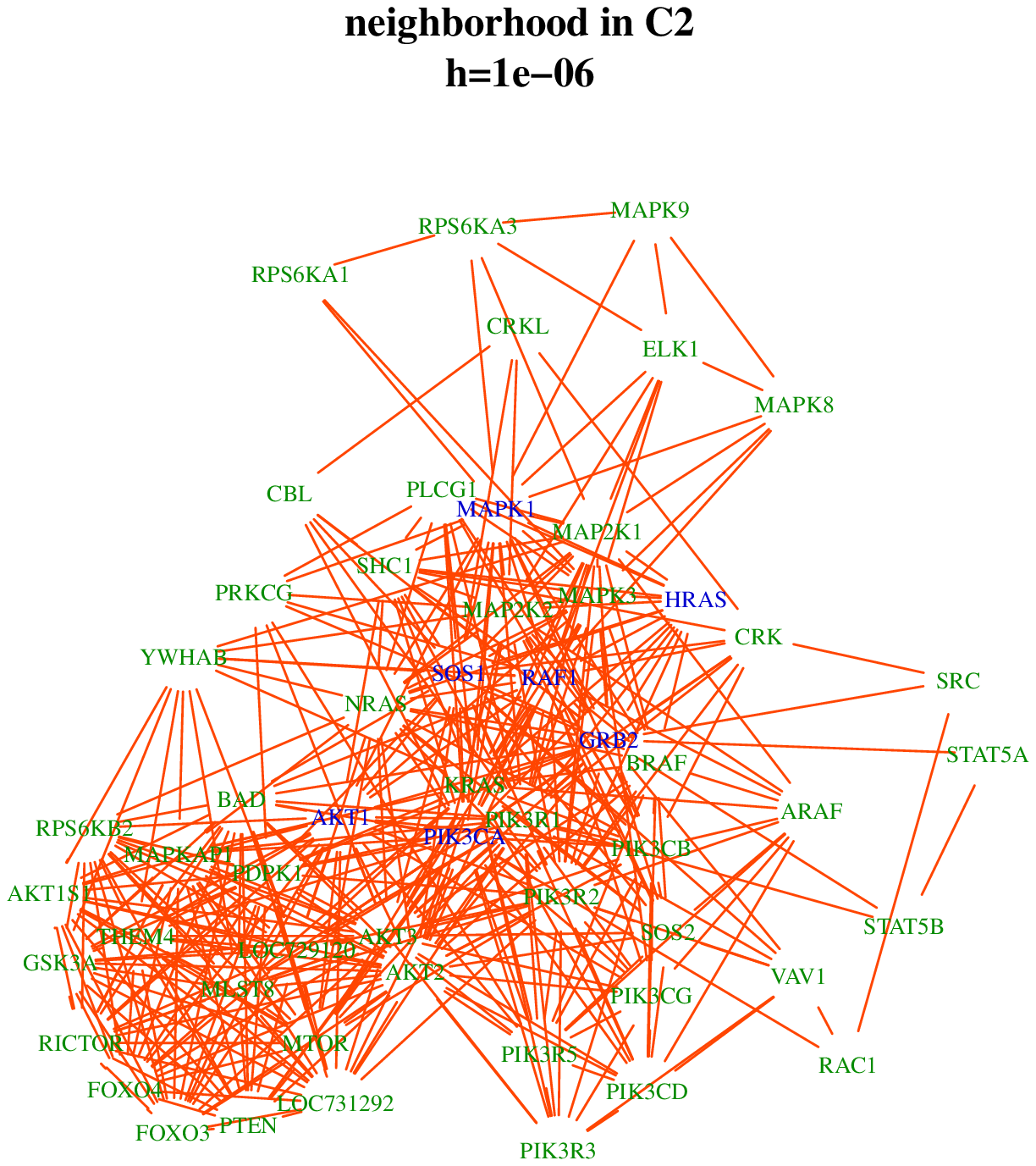}&
\includegraphics[width=7cm]{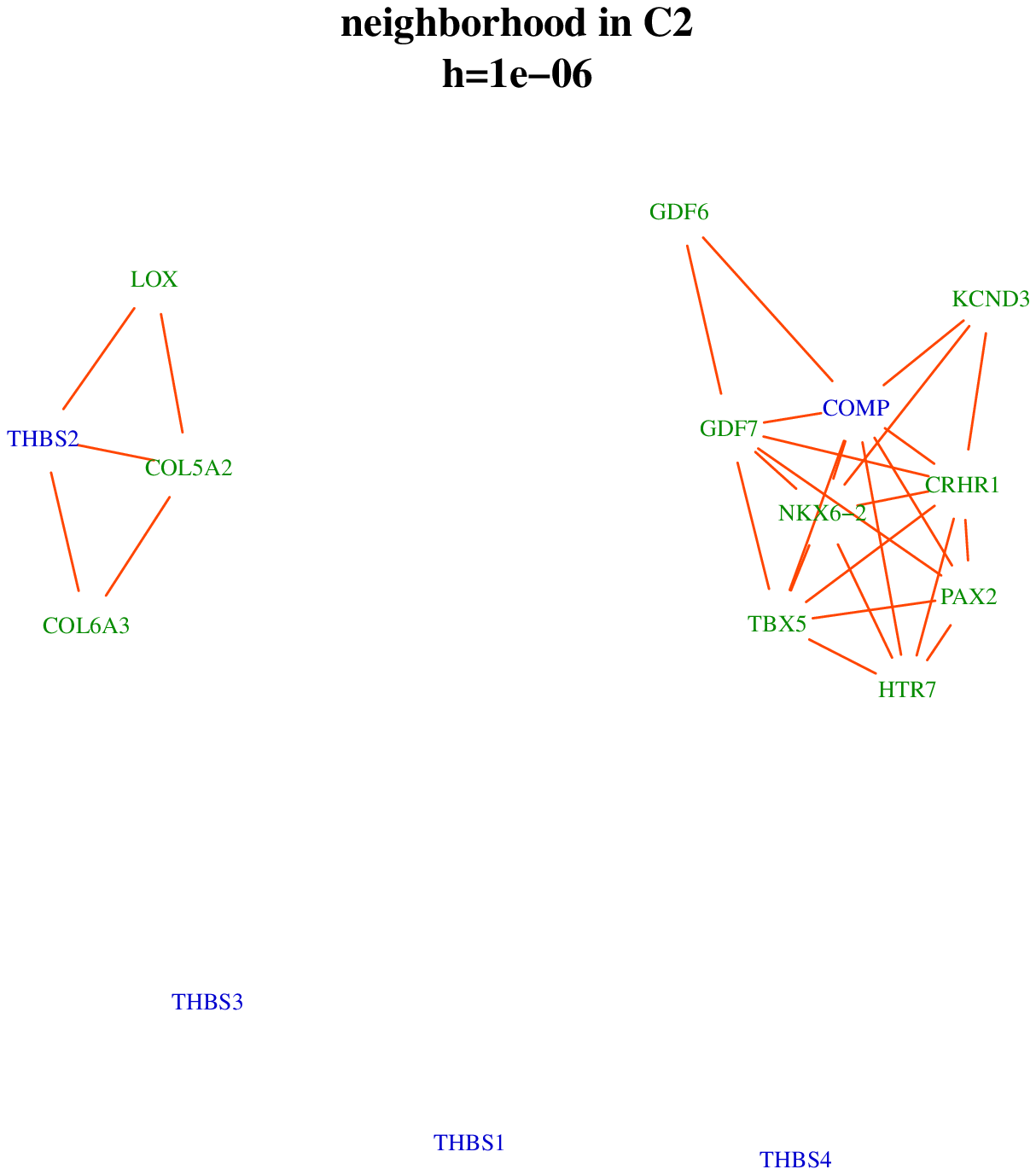}
\end{tabular}
} 
\caption{Neighbors at threshold $h=10^{-6}$ in \texttt{C2}. On the left
  panel, symbols 
AKT1, GRB2, HRAS, MAPK1, PIK3CA, RAF1, SOS1 appear in blue, their
neighbors in the association graph at threshold $h=10^{-6}$ appear in
green; the right panel shows a similar plot with COMP, THBS1--4.
}
\label{fig:AKT_neigh}
\end{figure}
%

\subsection*{Figure 4 - Detected associations in
\texttt{C2\_Cancer}}
At threshold $10^{-5}$ 946 significant associations were detected in
cancer related genesets of \texttt{C2}. For each association a point
represents the number of symbols and the number of genesets.
%
\begin{figure}[!ht]
\centerline{
\includegraphics[width=10cm]{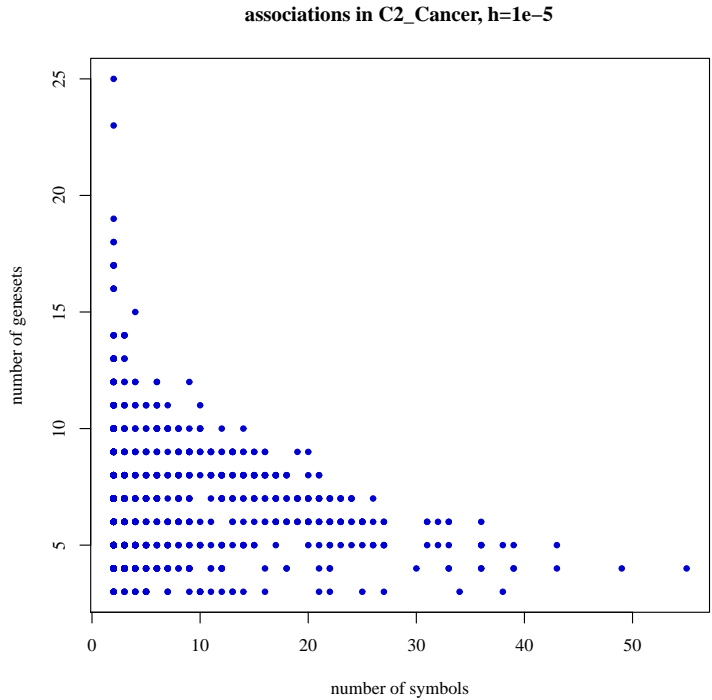}
} 
\caption{Detected associations in \texttt{C2\_Cancer}: 924 associations
  detected at threshold $10^{-5}$. For each association, the numbers of
  symbols (x-axis) and genesets (y-axis) involved are plotted.
}
\label{fig:Cliques_Cancer}
\end{figure}

\subsection*{Figure 5 - Hierarchical clustering of a group of symbols}
This Figure shows a hierarchical
clustering obtained through the single link algorithm in
\texttt{C2\_K}: 
the clusters
match known PPIs, and groups of symbols with the strongest
biochemical relation are correctly identified as homogeneous clusters. 
%
\begin{figure}[!ht]
\centerline{
\includegraphics[width=16cm]{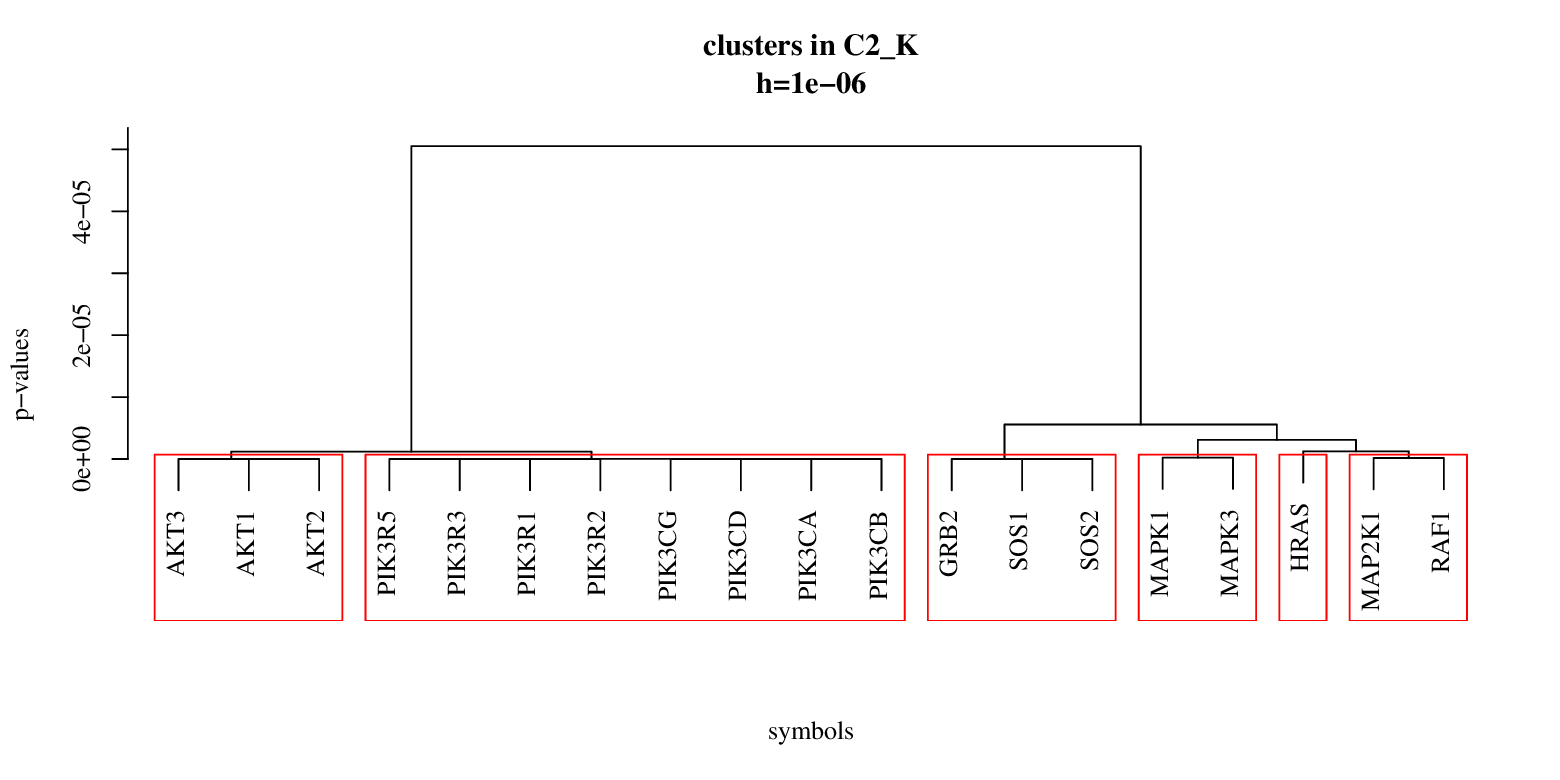}
}
\caption{Hierarchical clustering of a group of symbols 
in \texttt{C2\_K}.}
\label{fig:clusternet}
\end{figure}



\section*{Tables}
\subsection*{Table 1 - Neighbors at different thresholds in
\texttt{C2}}

For symbols AKT1, GRB2, HRAS, MAPK1, PIK3CA, RAF1, SOS1,
COMP, THBS1--4 (cf. Results section), and database
\texttt{C2}, the table gives
their frequency, and the number of neighbors at
thresholds $h=10^{-2},\ldots,10^{-10}$. \par \mbox{}
    \par
    \mbox{
      \begin{tabular}{|l||c|ccccccccc|}
        \hline 
symbols&frequency&\multicolumn{9}{|c|}{neighbors at threshold $h$}\\
&&$10^{-2}$&$10^{-3}$&$10^{-4}$&$10^{-5}$&$10^{-6}$&$10^{-7}$&$10^{-8}$&
$10^{-9}$&$10^{-10}$\\\hline
AKT1&    221& 70& 53& 45& 31& 28& 22& 16& 13&  8\\
GRB2&    216& 89& 63& 51& 38& 28& 23& 19& 16& 14\\
HRAS&    224& 76& 49& 33& 21& 15& 14& 14& 13& 12\\
MAPK1&   263& 80& 52& 41& 28& 21& 13& 12& 11& 11\\
PIK3CA&  235& 77& 60& 49& 41& 34& 28& 24& 20& 18\\
RAF1&    186&106& 65& 46& 37& 28& 21& 17& 14& 12\\
SOS1&    191& 92& 61& 44& 31& 26& 20& 19& 16& 14\\\hline
COMP&    35& 724& 366& 147& 59& 8& 1& 1& 0& 0\\
THBS1&  163&   7&   0&   0&  0& 0& 0& 0& 0& 0\\
THBS2&   86&  92&  27&   9&  6& 3& 2& 2& 0& 0\\
THBS3&   23& 213&  26&   6&  2& 0& 0& 0& 0& 0\\
THBS4&   37& 100&  13&   3&  0& 0& 0& 0& 0& 0\\\hline
      \end{tabular}
      }

\subsection*{Table 2 - Detected associations}

For \texttt{C2} and the 5 selections given as example, 
the number of genesets and the threshold at
which the detection was made are given as columns 2 and 3. 
The list
of detected associations can be sorted on numeric criteria. For instance:
number of symbols per association smaller than 10 (column 4), number
of genesets larger than 2 (column 5) or both (column 6). 
The lists are given as additional text files.\par \mbox{}
    \par
    \mbox{
      \begin{tabular}{|l||c|ccccc||}
        \hline 
&&\multicolumn{5}{|c|}{detected associations}\\
selection&genesets&threshold&all&pr$<10$&
gs$>2$&both\\\hline
\texttt{C2}C2&  4722&$10^{-15}$&  689& 500 &  689&  500\\ 
\texttt{C2\_Blast}&  57&$0.05$&  265&  249&  49&  49\\ 
\texttt{C2\_Breast}& 159&$10^{-2}$&  1337& 1044& 763& 628\\
\texttt{C2\_Cancer}& 948&$10^{-5}$& 924&  828&  924& 828\\
\texttt{C2\_K}&      186&$10^{-3}$& 501&  404&  334& 288\\
\texttt{C2\_Lymph}&  107&$10^{-2}$&  364&  325&  139& 132\\\hline
      \end{tabular}
      }


\section*{Additional Files}
Additional material has been provided as a compressed directory
available online:
\begin{verbatim}
http://ljk.imag.fr/membres/Bernard.Ycart/publis/spa.tgz
\end{verbatim}
It contains:
\begin{enumerate}
\item one subdirectory \texttt{C2}: 4722
  genesets, given as text files  
\item one R script file \texttt{spa.r}: the R functions implementing the
  method described here
\item one pdf file \texttt{spa\_manual.pdf}: a user manual for the R
  functions
\item
six text files \texttt{Cliques\_xyz.txt}: lists of associations
detected on the six
databases \texttt{xyz} (cf. Table 2).
\end{enumerate}

%

\end{bmcformat}
\end{document}